\documentclass[aps,prc,groupedaddress,showpacs]{revtex4}
\usepackage{graphicx,bm,color}

\newcommand{\slashi}[1]{\rlap{\sl/}#1}

\begin{document}

\title{Bakamjian-Thomas mass operator for the few-nucleon system from
  chiral dynamics}

\author{L. Girlanda}
\email[]{girlanda@pi.infn.it}
\affiliation{INFN, Sez. di Pisa, Largo Bruno Pontecorvo, 56127 Pisa, Italy}

\author{W. H. Klink}
\email[]{william-klink@uiowa.edu}
\affiliation{Department of Physics and Astronomy and Department of
  Mathematics, University of Iowa, Iowa City, Iowa 52242}

\author{M. Viviani}
\email[]{viviani@pi.infn.it}
\affiliation{INFN, Sez. di Pisa, Largo Bruno Pontecorvo, 56127 Pisa, Italy}

\date{\today}

\begin{abstract}
We  present an exploratory study consisting in the  formulation of a
relativistic quantum mechanics to describe the few-nucleon system
at low energy, starting from the quantum field theoretical chiral
Lagrangian involving pions and nucleons.
To this aim we construct a Bakamjian-Thomas mass operator and perform a
truncation of the Fock space which respects at each stage the relativistic
covariance. Such truncation is justified, at sufficiently low energy, in
the framework of a systematic chiral expansion. As an illustration we
discuss the bound state observables and low-energy phaseshifts of the
nucleon-nucleon and 
pion-nucleon scattering at the leading order of our scheme.

\end{abstract}


\maketitle

\section{Introduction}

The quantitative understanding of the dynamics of few-nucleon
systems, has reached such  
a high degree of accuracy that the need of 
including relativistic effects has been put forward
\cite{coester86,carlson93,forest95}. 
Starting from a non-relativistic setting, for a given 
 set of particles, it is possible to put constraints on the mutual
 interactions in order to fulfill the requirements of relativity. This
 problem has been posed in full generality by Dirac in 1949
 \cite{dirac}, and developed among others by Krajcik and Foldy \cite{krajcik}.
The relativistic corrections to the potential (``drift potentials'') can
take the 
form of multinucleon forces and are currently being investigated
quantitatively \cite{robilotta2006}; in some cases they are being used
to help in 
 the resolution of persisting puzzles in the few-nucleon physics,
like the neutron-deuteron $A_y$ problem \cite{miller}.

Since the pioneering work of Weinberg and
collaborators \cite{weinberg,vankolck}, the nuclear interaction
potential can nowadays be
thought of as derived from a (relativistic) quantum field theory based on the
chiral symmetry, which is the low-energy effective
theory of QCD (see Ref.~\cite{epelbaumreview} for a recent
review). A systematic perturbative expansion in powers of
$p/\Lambda_{\mathrm{H}}$ (Chiral Perturbation Theory, ChPT),  the ratio of
typical momenta divided by the 
typical hadronic scale $\Lambda_{\mathrm{H}}\sim$1~GeV, is formally
possible, whose convergence properties have to be checked case by case
in actual calculations.  
For power counting purposes a non-relativistic reduction is 
usually performed (heavy baryon expansion, HBChPT). This happens at
two levels: in the calculation of Feynman amplitudes and in the
definition of the effective potential, formulated by Weinberg in the
framework of old-fashioned (time-ordered) perturbation theory. While
the first step is not strictly necessary (a
relativistic covariant scheme for the calculation of
the two-pion exchange $NN$ potential exists \cite{robilotta} which relies on
the covariant version of baryon ChPT formulated by Becher and
Leutwyler \cite{beleut}), the second step seems unavoidable. Thus
relativistic and chiral corrections get 
intertwined, despite the different character of the corresponding
symmetries: Lorentz invariance is an exact symmetry of Nature, whereas 
chiral symmetry, although useful as organizing principle, is anyhow
approximate.  One might therefore want 
to 
treat relativity exactly. The question then arises whether it is
possible to set up a 
framework in which relativity is built in from the beginning, and corrections
come only from dynamics (in our case from higher chiral orders). To
the best of our knowledge this is not done 
   in current ChPT approaches to few-nucleon systems. By
separating the two expansions, one would also be able to compare relativistic
and chiral corrections, and check quantitatively the effectiveness of the
commonly adopted combined chiral and non-relativistic
expansion. Indeed such a test is only possible, in our opinion, by
performing a complete relativistic calculation, within a given
framework, and comparing to its non-relativistic limit.
Moreover a fully relativistic setting would allow to describe particle
production, contrary to non-relativistic quantum-mechanical treatments. 

The approach we take in the present paper is therefore to describe the
few-nucleon systems with a relativistic 
coupled-channel wave
equation, with simple transformation properties under the Lorentz
symmetry, so that the behaviour of e.g. a $N-N$ subsystem in relative
motion with respect to other nucleons, as demanded by relativity, is
easily accounted for.
In this perspective we consider the proposal of Ref.~\cite{klink00c} (see
also Ref.~\cite{fuda} for a similar point of view)
consisting in a Bakamjian-Thomas construction formulated in the
point-form of relativistic dynamics. This proposal has already been
implemented for bound state problems in the framework of chiral quark models
\cite{graz}. Compared  to the above references we focus on nucleons instead of
quarks and adopt a systematic ChPT-inspired point of view, instead of relying
on specific models.
We  apply the above mentioned framework to
 nucleons interacting with pions according to the dictates of chiral
 symmetry, using the vertices of the chiral Lagrangian at the lowest
 order, and restrict ourselves to the 1 nucleon sector ($\pi-N$
 scattering) and to the 2  nucleon sector (both the bound and
 scattering states of $N-N$). By adjusting the four
 low-energy constants which appear at the leading order, we obtain a
 reasonable unified 
 description of these systems at low energy. However, 
 the emphasis in the present paper is not 
 on the accuracy of the description (limited to a leading
 order treatment), but rather on the presentation  of a relativistic
 formalism which can naturally extend beyond the threshold of pion
 production.  

 The plan of the paper is as follows: a brief review of the
 adopted framework is given in the next section, in which the mass
 operator acting on the truncated Fock space is constructed from a
 Lagrangian density; in Section~\ref{sec:eft} we give a power-counting argument to justify the
 truncation of the Fock 
 space in the low-energy expansion in the case in which the
 interactions are restricted by chiral symmetry.
Section~\ref{sec:pin} contains the application to the 1-nucleon
sector, which yields the nucleon mass renormalization and a
Lippman-Schwinger type equation for the $\pi-N$ scattering. We adjust
the low-energy constant appearing at this order to reproduce the
$S$ wave scattering lengths; the pion-nucleon axial coupling $g_A$ is
fixed by the 
peripheral $NN$ phaseshifts, obtaining a 
reasonable 
description of the phaseshifts at small laboratory momentum (less than
50~MeV). In Section~\ref{sec:nn} we consider the 2-nucleon sector. Two
more vertices (contact interactions) arise at the lowest order and we 
adjust the corresponding coupling constants to reproduce the $S$-wave
scattering lengths. The agreement with experimental phaseshifts is
again reasonable, up to center-of-mass kinetic energies of 100~MeV.  

\section{Mass operator from a Lagrangian density}
\subsection{Bakamjian-Thomas construction in the point-form}
The requirements of  relativity
 are given by the Poincar\'e commutation relations among the generators of the
group, written in terms of the particle coordinates. This translates into
specific constraints for the possible interactions 
to include in the generators. Dirac classified three different possibilities
each one associated with a particular spacelike hypersurface left invariant by
a subgroup of the Poincar\'e  group: for the instant-form the hypersurface is
the hyperplane $t=$const., for the point-form it is the hyperboloid $t^2 -
{\mathbf x}^2 = \tau^2$, for the light-front it is the hyperplane $t+z=0$. The
generators associated with these hypersurfaces are said to be ``kinematical'', and
do not contain interactions. In the case of the point-form the Lorentz
transformations are kinematical and are the same as in the free case. Only the
four momentum $P^\mu$ contains interactions, and the requirements of
relativity are simply written covariantly as
\begin{equation}
\left[ P^\mu, P^\nu \right] =0, \quad U_\Lambda P^\mu U_\Lambda^{-1} = \left(
  \Lambda^{-1} \right)^\mu_{\,\nu} P^\nu,
\end{equation}
where the generators of the Lorentz transformation $\Lambda$ are unaffected by
interactions.
The Bakamjian-Thomas construction \cite{bt,b} in the point-form consists in
the definition, starting from the non-interacting Poincar\'e generators,  of a
mass 
operator $M_0=\sqrt{P_0^\mu P_{0 \mu}}$ and a four-velocity operator $V^\mu$
such that $P_0^\mu = M_0 V^\mu$; one then adds the interactions only to the
mass operator $M=M_0 + M_I$, and reconstructs the interacting four-momentum as
$P^\mu=M V^\mu$.   
Poincar\'e commutation relations are then satisfied provided the interacting
mass operator is a Lorentz scalar which commutes with the four-velocity
$V^\mu$.
It is therefore particularly convenient to consider the  ``velocity states''
\cite{klink98,graz}: 
these are linear combinations of 
multiparticle momentum states which are eigenstates of the four-velocity
operator. They have the nice property that all the particles transform with
the same Wigner rotation under a Lorentz transformation. Starting with (non
interacting) 
$n$-particle states $| p_1 
\sigma_1, p_2 \sigma_2,...,p_n \sigma_n \rangle$ with individual
four-momentum $p_i$ and spin projection $\sigma_i$, one defines the
internal momenta by going, through a canonical boost $B_c(v)$, to the
center-of-mass rest frame,
\begin{equation}
k_i=B_c^{-1}(v) p_i.
\end{equation}
By definition $\sum {\mathbf k}_i = 0$. 
$B_c(v)$ is a rotationless boost which transforms the system from its rest
frame to total velocity ${\mathbf v} = \sum_i {\mathbf p_i} / \sum_i{\omega_{\mathbf
      k_i}}$, where $\omega_{\mathbf k}$ is the relativistic energy of a free
particle with three-momentum ${\mathbf k}$. A velocity state is obtained from a 
multiparticle
momentum state defined in its rest frame after a boost to overall
velocity $v$ by means of $B_c(v)$,
\begin{equation} \label{eq:velstates}
| v, {\mathbf k}_1, \sigma_1,...,{\mathbf k}_n, \sigma_n\rangle \equiv U_{B_c(v)} |
 k_1,\sigma_1,...,k_n,\sigma_n\rangle.   
\end{equation}
 With this definition, using the
Lorentz invariant normalization for the momentum states, the velocity
states are normalized as follows,
\begin{equation} \label{eq:normalization}
\langle v, {\mathbf k}_1,\sigma_1,...,{\mathbf k}_n,\sigma_n|v',
{\mathbf k'}_1,\sigma_1',...,{\mathbf k'}_n,\sigma_n'\rangle = (2 \pi)^{3 n}
\frac{\prod_{i=1}^n 2 \omega_{{\mathbf k_i}}
  \delta_{\sigma_i,\sigma_i'}}{(\sum_{i=1}^n \omega_{{\mathbf k_i}})^3} 
  v_0\delta^3({\mathbf v}-{\mathbf v'}) \prod_{i=1}^{n-1} \delta^3 ( {\mathbf
    k}_i   - {\mathbf k'}_i)  .
\end{equation}

A Bakamjian-Thomas construction in the point-form is thus
accomplished in practice by defining the mass operator to be diagonal
in the four velocity \cite{klink00c}, which is conveniently expressed in terms
of velocity states as
\begin{equation} \label{eq:MIdef}
\langle v', {\mathbf k}_i',\sigma_i'|M_I| v,{\mathbf k}_i,\sigma_i\rangle = \langle v',
{\mathbf k}_i',\sigma_i'|{\cal H}(0)| v,{\mathbf k}_i,\sigma_i\rangle v_0 (2 \pi)^3
\delta^3({\mathbf v'} - {\mathbf v}) \frac{f( m , m')}{\sqrt{m^3 m'^3}},
\end{equation}
$m$ and $m'$ being the  initial and final relativistic energies,
$m= \sum_i \omega_{\mathbf k_i}$, $m' = \sum_i \omega_{\mathbf k_i'}$.
In this equation ${\cal H}(x)$ is the density of the interaction
Hamiltonian, which is a Lorentz scalar. Therefore, the above
definition is frame-independent, in view of Eq.~(\ref{eq:velstates}),
and in principle any rotationally-invariant combination of the momenta
${\bf k_i}$ and ${\bf k_i'}$ would be allowed to appear.
 ${\cal H}(x)$ will be taken as
a sum of vertices constructed as local products of field operators. 
The structure function $f$ is introduced
in order to compensate for the neglect of the off-diagonal terms in
the four velocity, and to regulate the integrals as well.   From its
definition the structure function $f$ is dimensionless. From now on we will
take for $f$ a real symmetric function of its arguments, further specified as a
Gaussian function centered around zero with cutoff 
$\Lambda$, times an additional cutoff function $\xi$ of the relativistic
invariants, which may be needed in order to regulate the integrals, 
\begin{equation} \label{eq:cutoff}
f(m,m')  =  {\mathrm{exp}} \left[-\frac{( m - m')^2}{2
    \Lambda^2} \right] \xi.
\end{equation}
The cutoff $\Lambda$, within the effective  theory implementation of the
present approach, should be understood as the short distance scale at
which new degrees of freedom start to become relevant. 
One such structure function is understood for each vertex of the
interaction Hamiltonian, and the requirement that physics at low-energy is
independent of the cutoff (provided it is large enough) should fix the
running of the coupling constants with $\Lambda$. 
Compared to
Ref.~\cite{klink00c} we have introduced in Eq.~(\ref{eq:MIdef}) a different normalization for the 
matrix elements of the interacting mass operator, in order to properly match, in the case when  $v = v'$ and $m=m'$,  with the quantum field theoretical  result \cite{fubini} 
\begin{equation}
P_\mu^{\mathrm{int}}=\int d^4 x \frac{\partial F(x)}{\partial x^\mu} \delta(F(x) - \tau^2) {\cal H^{\mathrm{int}}}(x),
\end{equation} 
where in the point-form $F(x)=x^2$ and the factor $m^3$ appears as a
Jacobian in passing from the overall momentum 
conserving $\delta$ function to the velocity conserving $\delta$ function.

\subsection{A simple toy-model}
\begin{figure}
\centerline{\includegraphics[width=8cm,angle=0]{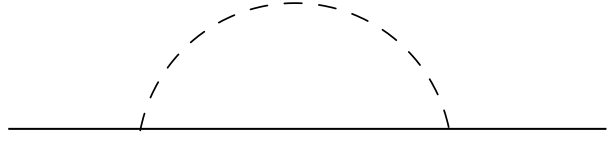}}
\caption{\label{fig:massren} Diagram contributing to the nucleon mass renormalization.}
\end{figure}

As an illustration of the general setting, we start by examining  the simple
example  of a scalar nucleon field $\Psi$ interacting with a pion field 
$\phi$, where the interactions are provided by a
Hamiltonian density 
of the form ${\cal H}(x)=g \Psi^\dagger(x) \Psi(x) \phi(x)$. (Here and in the
following all
products of field operators are understood as normally ordered.) Creation
of nucleon-antinucleon pairs is neglected and a 
truncation of the Fock space to a given maximum number of pions is
considered from the beginnning. In the 1-nucleon sector, truncating
the states containing two or more pions, the mass
operator takes the form 
\begin{equation}
M=\left( \begin{array}{cc}
m_N + \delta^{\mathrm{ren}}_1 & g K \\
g K^\dagger & D_{1+1} \end{array} \right),
\end{equation}
where $m_N$ is the physical nucleon mass, and $D_{1+1}$ is the relativistic
1-nucleon + 1-pion free particle energy. The counterterm
$\delta^{\mathrm{ren}}_1$ is needed for the mass renormalization. Due to the
form of ${\cal H}(x)$, the 
interactions show up as off-diagonal entries in the mass operator.
The nucleon mass renormalization and pion-nucleon scattering are
described as eigenvalue-eigenvector problems for this mass operator. 
For instance, for the eigenvalue $m_N$, the physical nucleon mass,
one finds an equation for the counterterm
\begin{equation}
\delta^{\mathrm{ren}}_1 =  g^2  K^\dagger ( D_{1+1} - m_N )^{-1} K  , 
\end{equation}
with $D_{1+1} = \omega_{\mathbf k} + \omega^\pi_{\mathbf k}$, having defined
\begin{equation}
\omega_{\mathbf k} \equiv \sqrt{m_N^2 + {\mathbf k}^2}, \quad \omega^\pi_{\mathbf k} =
\sqrt{M_\pi^2 + {\mathbf k}^2},
\end{equation}
and the operator $K$, which 
    connects 2-particle to 1-particle states,
\begin{eqnarray}
\langle v', {\mathbf k},-{\mathbf k} | g K | v,{\mathbf 0} \rangle &\equiv & g
\frac{f^{(1)} (m , m')}{\sqrt{m^3 m'^{3}}}
\langle v', {\mathbf k}, - {\mathbf k} | \Psi^\dagger (0) \Psi(0) \phi (0) | v,{\mathbf 0} \rangle  v_0 (2
\pi)^3 \delta^3 ( {\mathbf v} - {\mathbf v'})  \nonumber \\
&=& g \frac{f^{(1)} (m_N, \omega_{\mathbf k} + \omega^\pi_{\mathbf k})}{\sqrt{m_N^3 (\omega_{\mathbf k} + \omega^\pi_{\mathbf k})^3}} v_0 (2 \pi)^3 \delta^3( {\mathbf
  v} - {\mathbf v'}).
\end{eqnarray}
The superscript $^{(1)}$ refers to the sector of the Fock space with
baryon number 1. The mass operator commutes with the baryon number,
and there is the freedom to choose a different structure function $f$
for each sector of the Fock space.
Taking the expectation value of the above equation between 1-nucleon
states and inserting a complete set of velocity states in the subspace of
1-nucleon + 1-pion states one arrives at the nucleon mass
renormalization due to the ``pion cloud'',
\begin{equation}
\label{eq:massren}
\delta^{\mathrm{ren}}_1 =  \frac{g^2 }{2 m_N} \int \frac{d^3 {\mathbf k}}{(2 \pi)^3} \frac{1}{4 \omega_{\mathbf k} \omega^\pi_{\mathbf k}} \frac{ |f^{(1)} (m_N,
  \omega_{\mathbf k}
  +\omega^\pi_{\mathbf k}  )|^2}{ \omega_{\mathbf k} + \omega^\pi_{\mathbf k}-m_N } .
\end{equation}
Eq.~(\ref{eq:massren})  determines the counterterm $ \delta^{\mathrm{ren}}_1$ for
each choice of the cutoff $\Lambda$ and coupling constant $g$, and 
corresponds diagramatically  to the process shown in
Fig.~\ref{fig:massren}. 

In the 2-nucleon sector, an analogous equation describes the deuteron,
\begin{equation} \label{eq:eigenstates}
\left( D_2 + \delta^{\mathrm{ren}}_2 \right) \phi_2^D + g^2 K^\dagger ( m_D - D_{2+1})^{-1} K \phi_2^D = m_D
\phi_2^D,
\end{equation}
where $\phi_2^D$ is a state vector in the subspace of 2-nucleon
states, and the operators $D_2$ and $D_{2+1}$ are respectively the
relativistic 2-nucleon and 2-nucleon + 1-pion energy.
As in the 1-nucleon sector, a counterterm $ \delta^{\mathrm{ren}}_2$ is
introduced in the corresponding diagonal element of the mass operator,
in order to 
properly renormalize the 2-particle states. 
The kernel $K$ connects 2-particle to 3-particle states and has matrix
elements between velocity states
\begin{eqnarray}
&&\langle v', {\mathbf k_1}, {\mathbf k_2}, -{\mathbf k_1} - {\mathbf k_2} | g K | v, {\mathbf q},
-{\mathbf q} \rangle = g \frac{f^{(2)} (
 m , m')}{\sqrt{m^3 m'^3}} v_0 (2 \pi)^3
\delta^3( {\mathbf v} - {\mathbf v'})  \\
&& \times (2 \pi)^3\left[ 2 \omega_{\mathbf k_1} \delta^3( {\mathbf k_1} -
  {\mathbf q} ) + 2 \omega_{\mathbf k_1} \delta^3 ( {\mathbf k_1} + {\mathbf q}) + 2
  \omega_{\mathbf k_2}
  \delta^3 ({\mathbf k_2} - {\mathbf q}) + 2 \omega_{\mathbf k_2} \delta^3({\mathbf k_2} +
        {\mathbf q}) \right], \nonumber 
\end{eqnarray}
with $m=2 \omega_{\mathbf q}$ and $m'=\omega_{\mathbf k_1} +
\omega_{\mathbf k_2} + \omega^\pi_{{\mathbf k}_1 + {\mathbf k}_2}$.
For what concerns the covariance properties, the structure function
$f^{(2)}$ for the 2-nucleon sector can be chosen different from
$f^{(1)}$. However, as we will see in subsection~\ref{sec:ren}, in
order to have a consistent renormalization we have to choose
$f^{(1)}=f^{(2)}$.  
By left-multiplying Eq.~(\ref{eq:eigenstates}) with the bra $\langle
v, {\mathbf k}, -{\mathbf k} |$ representing  a 2-nucleon state with four velocity $v$
and relative momentum (in the center-of-mass system) $2 {\mathbf k}$, one arrives, after insertion of a
complete set of states in the subspace of 2-nucleon + 1-pion states,
to an eigenvalue wave equation for the center-of-mass wave function $\phi_2^D({\mathbf
  k}) = \langle v=(1,{\mathbf 0}), {\mathbf k}, -{\mathbf k} | \phi_2^D\rangle$. Using Bose symmetry
[$\phi_2^D({\mathbf k}) = 
\phi_2^D(-{\mathbf k})$], the bound state equation becomes
\begin{equation} \label{eq:eigenwaves}
\left( 2 \omega_{\mathbf k} + \delta^{\mathrm{ren}}_2 ({\mathbf k})
\right)\phi_2^D({\mathbf k}) + 2 \omega_{\mathbf k} A({\mathbf k}) \phi_2^D
({\mathbf k}) 
+ \int \!\!\frac{d^3 {\mathbf q}}{(2 \pi)^3} B({\mathbf k}, {\mathbf q})
\phi_2^D({\mathbf q}) = m_D \phi_2^D({\mathbf k}),
\end{equation}
with
\begin{equation} \label{eq:functionA}
A({\mathbf k}) =    \int  \!\!\frac{d^3 {\mathbf q}}{(2 \pi)^3}
\left\{ \frac{g^2}{16 \omega_{\mathbf k}^2 \omega_{\mathbf q} \omega^\pi_{{\mathbf k} + {\mathbf
      q}}} \frac{ \left| f^{(2)} ( 2  \omega_{\mathbf k} ,  \omega_{\mathbf k}  + \omega_{\mathbf q} +
  \omega^\pi_{{\mathbf k} + {\mathbf q}} ) \right|^2}{m_D - \omega_{\mathbf k} -
  \omega_{\mathbf q} - \omega^\pi_{{\mathbf k}+{\mathbf q}} } + {\mathbf q}
\leftrightarrow - {\mathbf q} \right\}
\end{equation}
and
\begin{equation}
\label{eq:functionB}
B({\mathbf k},{\mathbf q}) 
=   \frac{ g^2 }{8 \omega^\pi_{{\mathbf k} +
    {\mathbf q}} \sqrt{\omega_{\mathbf q} \omega_{\mathbf k}^3} }  \frac{ f^{(2)}
  ( 2 \omega_{\mathbf q}, \omega_{\mathbf q} + \omega_{\mathbf k} + 
    \omega^\pi_{{\mathbf k}+{\mathbf q}}) f^{(2)} (\omega_{\mathbf q}+ \omega_{\mathbf k}  +
    \omega^\pi_{{\mathbf k}+{\mathbf q}}, 2  \omega_{\mathbf k})}{m_D - \omega_{\mathbf k} -
  \omega_{\mathbf q} - \omega^\pi_{{\mathbf k}+{\mathbf q}} }+ {\mathbf q}
\leftrightarrow - {\mathbf q}.
\end{equation}
\begin{figure}
\centerline{\includegraphics[width=8cm,angle=0]{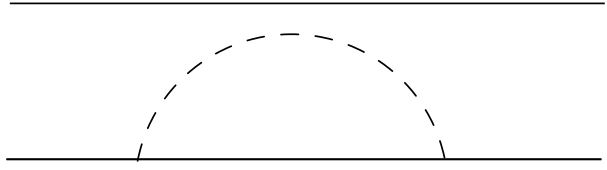}}
\caption{Diagram corresponding to the disconnected kernel of the two-particle
  scattering. \label{fig:2massren}}
\end{figure}
The  term  proportional to
$A({\mathbf k})$ represents a  wave function renormalization of the
two-nucleon state: it describes diagrams in which the nucleon lines
are disconnected and dressed with pion loops, cfr. Fig.~\ref{fig:2massren}.

It is possible to resum such diagrams and obtain a Lippmann-Schwinger
equation for a reduced (renormalized) amplitude containing only a connected
kernel.
Alternatively, one can choose the counterterm $\delta^{\mathrm{ren}}_2$ so as
to cancel the disconnected kernel,
$\delta^{\mathrm{ren}}_2({\mathbf k})=- 2 \omega_{\mathbf k} A ({\mathbf k})$. 

Correspondingly, the $NN$ scattering is described by the
Lippmann-Schwinger equation,
\begin{equation}
{ \Psi_2} = \phi_2 + g^2 \left. \left[ \sqrt{s} - D_2  \right]^{-1}
K^\dagger \left[ \sqrt{s} - D_{2+1} \right]^{-1} K
\right|_{\mathrm{conn}} { \Psi_2}, 
\end{equation}
where $\sqrt{s}=E + i \epsilon$ is the scattering energy, and
$\phi_2$ is an eigenstate of the free mass operator in the
two-particle subspace. The above equation can
be equivalently written in terms of the scattering amplitude, $T({\mathbf
  q}, {\mathbf k})$ defined as
\begin{equation}
T({\mathbf q},{\mathbf k}) = g^2 \langle v,{\mathbf q},-{\mathbf q} |\left. K^\dagger \left[\sqrt{s} -
  D_{2+1} \right]^{-1} K \right|_{\mathrm{conn}}| \Psi_2 \rangle_{\mathbf k},
\end{equation}
where ${\mathbf k}$ denotes the incident three-momentum of the interacting state
$\Psi_2$.
Inserting a complete set of velocity states, and  omitting the
three-delta over the velocities (such factors will
always be present because they appear in every interaction
vertex)  the LS equation takes the form
\begin{equation}
T({\mathbf q},{\mathbf k}) = V({\mathbf q},{\mathbf k}) + \int  \omega_{\mathbf
  p}\frac{d^3 {\mathbf p}}{(2 \pi)^3} \frac{V({\mathbf q},{\mathbf p}) T({\mathbf
  p},{\mathbf k})}{\sqrt{s} - 2 \omega_{\mathbf p}  + i \epsilon},
\end{equation}
where the potential,
\begin{equation} \label{eq:Vqk}
V({\mathbf q},{\mathbf k}) = g^2 \langle v,{\mathbf q},-{\mathbf q} |
\left.  K^\dagger \left[\sqrt{s} - 
  D_{2+1} \right]^{-1} K\right|_{\mathrm{conn}} |v ,{\mathbf k} ,
-{\mathbf k}\rangle = B({\mathbf q},{\mathbf k}),
\end{equation}
consists only of the connected kernel $B$ in Eq.~(\ref{eq:functionB}),
with the substitution $m_D \to \sqrt{s}$.

\subsection{Consistency of the renormalization procedure} \label{sec:ren}
The renormalization of the 2-nucleon lines describing $NN$ scattering,
realized by the choice of 
the counterterm  $\delta_2^{\mathrm{ren}}
({\bf k}) = -2 \omega_{\mathrm{\bf k}} A({\mathbf k})$, and of the
1-nucleon line, Eq.~(\ref{eq:massren}), correspond to the same
physical processes, as can be seen by comparing
Figs.~\ref{fig:massren} and~\ref{fig:2massren}. Physical
considerations would require that, when the two nucleons are far apart
and at rest,  their energies should be renormalized as their respective
masses. This implies the condition
\begin{equation}
\delta_2^{\mathrm{ren}}({\bf 0}) = 2 \delta^{\mathrm{ren}}_1,
\end{equation}
which can be regarded as the manifestation of the cluster
decomposition principle in the simple case of two particles. The
equation to fulfill is therefore
\begin{equation}
-2 m_N \int  \!\!\frac{d^3 {\mathbf q}}{(2 \pi)^3}
\left\{ \frac{g^2}{8 m_N^2 \omega_{\mathbf q} \omega^\pi_{ {\mathbf
      q}}} \frac{ \left| f^{(2)} ( 2  m_N ,  m_N  + \omega_{\mathbf q} +
  \omega^\pi_{ {\mathbf q}} ) \right|^2}{m_N - 
  \omega_{\mathbf q} - \omega^\pi_{{\mathbf q}} }  \right\}
=2 
\frac{g^2 }{2 m_N} \int \frac{d^3 {\mathbf q}}{(2 \pi)^3} \frac{1}{4 \omega_{\mathbf q} \omega^\pi_{\mathbf q}} \frac{ |f^{(1)} (m_N,
  \omega_{\mathbf q}
  +\omega^\pi_{\mathbf q}  )|^2}{ \omega_{\mathbf q} +
  \omega^\pi_{\mathbf q}-m_N },
\end{equation}
where we have replaced in Eq.~(\ref{eq:functionA}) $m_D$ by $\sqrt{s}=2 m_N$,
since we are considering the case of two widely separated nucleons at 
rest. It is seen that our choice of the structure function,
 $f^{(1)}= f^{(2)}=f$ depending on $m-m'$ as in Eq.~(\ref{eq:cutoff}),
independently of 
the baryon number sector, satisfies the
requirement of a consistent renormalization procedure. Notice that
this would not happen had we chosen the original formulation of
Ref.~\cite{klink00c}: the crucial point was the inclusion of a
different normalization for the matrix elements of the interacting
mass operator, Eq.~(\ref{eq:MIdef}), which in turn was dictated by a
proper matching to the quantum field theory. The cluster decomposition
principle, satisfied by local quantum field theories, could in general
be violated by a truncation of the full quantum field theory to a
relativistic quantum mechanics. In view of the above consideration, we
will from now on drop the superscripts and use the same structure
function $f$ for all sectors of the Fock space.

\section{Effective theory implementation} \label{sec:eft}
Having described  the general features of the construction of the
interacting mass operator from a vertex Lagrangian, we now proceed to
make full use of the constraints given by chiral symmetry. Most
importantly, the Goldstone theorem requires that the coupling between
pion and nucleons be of derivative type (suppressed at low energy),
and this will in turn provide a power-counting justification for the
truncation of the Fock space, since the creation of pions brings more
and more powers of momentum.
A lowest order interaction Lagrangian which respects chiral symmetry is 
\begin{equation}
{\cal L}_{\pi N} = -\frac{g_A}{2 F_\pi} \bar \psi  \gamma^\mu \gamma_5
\partial_\mu \pi \psi.
\end{equation}
In this expression $\psi$ is to be understood as an isospin spinor and
the pion field $\pi=\tau^a \pi^a$ as the Goldstone boson
SU(2) matrix, representing the coordinates of the coset space of chiral
symmetry breaking.
In addition to the pion-nucleon-nucleon vertex, at lowest order one
has to consider other vertices such as the Weinberg-Tomozawa or the
nucleon-nucleon contact interaction, which comes into two independent
operators, but let us for the time being concentrate on the above vertex.
It will contribute to the mass operator as off-diagonal
matrix elements. For instance, in the two-nucleon sector,
\begin{equation}
\left( \begin{array}{cccc}
\tilde D_2  & K & 0 & \\
K & D_{2+1} & K & \\
0 & K & D_{2+2} & \\
& & & \ddots
\end{array}
\right)
\left( \begin{array}{c} \phi_2 \\ \phi_{2+ 1}\\ \phi_{2+2} \\ \vdots
\end{array} \right) = \sqrt{s} 
\left( \begin{array}{c} \phi_2 \\ \phi_{2+1} \\ \phi_{2+2} \\ \vdots
\end{array} \right),
\end{equation}
where we have defined for ease of notation $\tilde D_2 = D_2 + \delta^{\mathrm{ren}}_2$.
Here  $K$ connects states containing $n$ and $n+1$ pions and is defined as in
Eq.~(\ref{eq:MIdef}) using ${\cal H}(x) = -{\cal L}_{\pi N}$. 
 The multi-particle
states $\phi_{2+n}$ are expanded in velocity 
   states which are 
normalized as in Eq.~(\ref{eq:normalization}). Counting three-momenta
as small parameters of order $O(p)$, the chiral power of the
phase space element $d_{2+n} \phi$, such that
\begin{equation}
{\mathbf 1} = \int d_{2+n} \phi | \phi_{2+n} \rangle \langle \phi_{2+n} | ,
\end{equation}
in the two-nucleon sector is 
\begin{equation}
d_{2+n} \phi \sim O(p^{2 n +3}).
\end{equation}
The formal solution of the above eigenvalue problem gives an equation
for the two-particle component of the state vector,
\begin{equation} \label{eq:continued2}
\left[ \sqrt{s} - \tilde D_2  - K \frac{1}{\sqrt{s} - D_{2+1} - K \frac{1}{\sqrt{s} - D_{2+2}
   - ...} K} K  \right] \phi_2 = 0.
\end{equation}
One can
Taylor-expand the denominator in the continued
fraction~(\ref{eq:continued2}), and compare the different operators
\begin{equation}
(\sqrt{s} - \tilde D_2 )   - K \frac{1}{\sqrt{s} - D_{2+1}} K  - K \left[
    \frac{1}{\sqrt{s} - D_{2+1}} \right]^2 K \frac{1}{\sqrt{s} - D_{2+2}} K K ..., 
\end{equation}
contracted between two-particle states. Since we are interested in kinematical configurations in which
$\sqrt{s} - \tilde D_2 \sim O(p)$, the first term goes like
\begin{equation}
\langle 2 | \sqrt{s} - \tilde D_2 | 2' \rangle \sim O(p^{-2});
\end{equation}
the second term, inserting a complete set of three-particle states,
\begin{equation}
\langle 2 | K \frac{1}{\sqrt{s} - D_{2+1}} K | 2'\rangle \sim O(p^0),
\end{equation}
where we have used the fact that
$\langle 2 | K | 2+1 \rangle \sim O(p^{-2})$,
while the third term, since $\langle 2 +1| K | 2+2 \rangle \sim O(p^{-4})$,
\begin{equation}
\langle 2 | K \left[
    \frac{1}{\sqrt{s} - D_{2+1}} \right]^2 K \frac{1}{\sqrt{s} - D_{2+2}
    } K K | 2'\rangle \sim O(p^2).
\end{equation}
The hierarchy is 
\begin{equation}
O(p^{-2}) + O(p^0) + O(p^2) + ...
\end{equation}
Therefore the inclusion of 4-particle states (2-nucleons +  2-pions) $\phi_{2+2}$ yields a
contribution suppressed by two orders in the chiral counting compared to the
one of 3-particle states (2-nucleons + 1-pion). This observation is the
basis to justify the truncation of the Fock space. It is possible to
convince oneself that the same mechanism applies for  
the most general vertex respecting chiral symmetry.
 In this paper we stick to the first order of the low-energy expansion.

\section{One nucleon sector} \label{sec:pin}
The vertices to be considered at the lowest order in the 1-nucleon
sector for the study of $\pi N$ scattering come from the Lagrangian
\begin{equation} \label{eq:Lpin}
{\cal L}_{\pi N} = -\frac{g_A}{2 F_\pi} \bar \psi  \gamma^\mu \gamma_5
\partial_\mu \pi \psi + \frac{i}{8 F_\pi^2} \bar \psi  \gamma^\mu
        [\pi, \partial_\mu \pi] \psi.
\end{equation}
These vertices are part of the leading chiral Lagrangian, expanded up to
terms quadratic in the pion field. From the argument presented above,
the neglected terms start to contribute at the next to leading order.
Only the $\pi N N $ vertex contributes to the nucleon mass
renormalization, which reads,
\begin{equation} \label{eq:chmassren}
\delta^{\mathrm{ren}}_1 =\frac{3 g_A^2}{4 F_\pi^2}  \int \frac{d^3{\mathbf k}}{(2 \pi)^3}  
\frac{{\mathbf k}^2}{4 \omega_{\mathbf k}
  \omega^\pi_{\mathbf k}} \frac{[\omega^\pi_{\mathbf k} + \omega_{\mathbf k} +
  m_N ]^2}{\omega_{\mathbf k} + m_N} \frac{ |f_{A} (
 m_N, \omega_{\mathbf k}  +\omega^\pi_{\mathbf k}  )|^2}{\omega_{\mathbf
    k} +   \omega^\pi_{\mathbf k} - m_N}.
\end{equation}

The subscript in $f$ denotes that the form factor is associated with
the axial $\pi N N$ vertex of the interaction Lagrangian. As in
HBChPT, the mass renormalization counts formally as $O(p^3)$. 

\begin{figure}
\centerline{\includegraphics[width=8cm,angle=0]{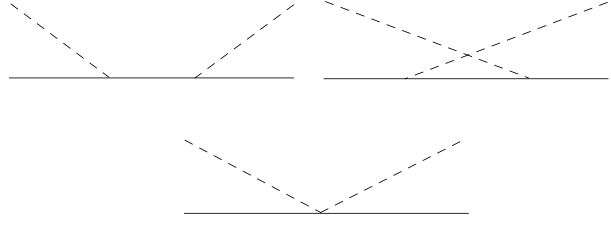}}
\caption{\label{fig:pindiagrams} Connected diagrams contributing to the $\pi N$
  scattering at the leading order, corresponding to the axial coupling (upper
  panel) and to the Weinberg-Tomozawa vertex (lower panel).}
\end{figure}

For the description of $\pi N$ scattering, one needs to take into
account the truncation up to 2 pions, because the $\pi N$ states are
connected through the leading vertex to pure $N$ states and to $\pi
\pi N$ states\footnote{One can show this along the same lines as done
  for the $NN$ scattering.}. This corresponds to the fact that the scattering
contains the direct nucleon pole and the crossed one, as shown in 
Fig.~\ref{fig:pindiagrams}, where the Weinberg-Tomozawa vertex is also included.

\begin{figure}
\centerline{\includegraphics[width=8cm,angle=0]{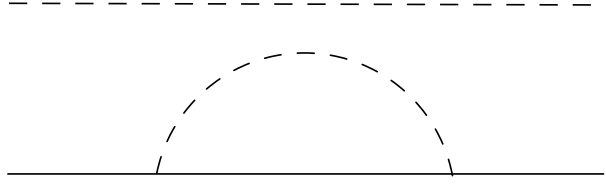}}
\caption{\label{fig:pindisconnected} Disconnected diagrams contributing to the
  $\pi N$   scattering at the leading order.}
\end{figure}

Consider then
the pion-nucleon scattering in the center of mass frame, with $\{ {\mathbf
  q}, r, i\}$ ($\{ {\mathbf
  k}, s, j\} $) the initial (final) nucleon momentum, spin and isospin
indices, and $\{  -{\mathbf q} ,a\}$ ($\{  -{\mathbf k} ,b\}$) the initial (final) pion
momentum and isospin index. 
The kernel is written as 
\begin{eqnarray}
  B^{r i a}_{s j b}({\mathbf q},{\mathbf k}) &=& \frac{g_A^2}{4 F_\pi^2}\frac{
f_A(\omega_{\mathbf q} + \omega^\pi_{\mathbf q}, m_N) f_A
(m_N,\omega_{\mathbf k} +
  \omega^\pi_{\mathbf k} )}{2 m_N \sqrt{( \omega_{\mathbf q} +
    \omega^\pi_{\mathbf q})^3 ( \omega_{\mathbf k} +
    \omega^\pi_{\mathbf k})^3} }  \frac{( \tau^a
    \tau^b)_{ij}}{\sqrt{s} - m_N -\delta^{\mathrm{ren}}_1}
  \nonumber \\
&& \times \bar
    u({\mathbf q},r) \slashi{q^\pi} \gamma_5 u({\mathbf 0},\sigma) \bar u({\mathbf 0},\sigma) \slashi{k^\pi} \gamma_5
  u({\mathbf k},s) \nonumber \\
&+&\frac{g_A^2}{4 F_\pi^2}  
  \frac{f_A(2  \omega_{\mathbf q},\omega_{{\mathbf q} +{\mathbf k}} + \omega_{\mathbf q} + \omega^\pi_{\mathbf k})
  f_A(\omega_{{\mathbf q}+{\mathbf k}} + \omega_{\mathbf k} + \omega^\pi_{\mathbf
    q},2 \omega_{\mathbf k} )}{2
    \omega_{{\mathbf q}+{\mathbf k}} \sqrt{( \omega_{\mathbf q} +
    \omega^\pi_{\mathbf q})^3 ( \omega_{\mathbf k} +
    \omega^\pi_{\mathbf k})^3 }  } 
\frac{   ( \tau^b \tau^a)_{ij}}{\sqrt{s} - \omega_{{\mathbf
      q}+{\mathbf k}} - \omega^\pi_{\mathbf
  q} - \omega^\pi_{\mathbf k}} \nonumber \\
&&\times \bar u ({\mathbf q},r) \slashi{k^\pi} \gamma_5 u (-{\mathbf
    q} - {\mathbf k},\sigma) \bar u  (-{\mathbf
    q} - {\mathbf k},\sigma) \slashi{q^\pi} \gamma_5 u ({\mathbf k},s) \nonumber \\
&+& \frac{i \epsilon^{abc} \tau^c_{ij}}{4 F_\pi^2} \frac{f_{WT}(\omega_{\mathbf k} +
\omega^\pi_{\mathbf k},  \omega_{\mathbf q} + \omega^\pi_{\mathbf q})}{ \sqrt{( \omega_{\mathbf q} +
    \omega^\pi_{\mathbf q})^3 ( \omega_{\mathbf k} +
    \omega^\pi_{\mathbf k})^3 }  } \bar u ({\mathbf q},r)
(\slashi{q^\pi} + \slashi{k^\pi}) u({\mathbf k},s),
\end{eqnarray}
where we have denoted by $q^\pi$ and $k^\pi$ the pion 4-momenta;
the first two terms come from the direct  and the
 crossed nucleon pole, while the third one from the Weinberg-Tomozawa
 vertex, with an associated structure function $f_{WT}$.
The Dirac 4-spinors are normalized so that $\sum_s u({\mathbf k},s) \bar u({\mathbf k},s)=
\slashi{k} + m_N$. The isospin structures are proportional to
 $\delta_{ab}$, (``isoscalar'')  that we will denote with a
 ``+'' superscript, and to $i
 \epsilon^{abc} \tau^c$, (``isovector'') that we will  denote
 with a ``-'' superscript. Analogously, the spin structures are
 proportional to $\delta_{r s}$  and to ${\mathbf q} \times {\mathbf k} \cdot
 {\bm \sigma}_{r s}$. The kernel can thus be written in the following
operatorial form,
\begin{eqnarray}
  B({\mathbf q}, {\mathbf k}) &=& \delta^{ab} \left[ g^+(q,k,\cos \theta) + i
  {\mathbf q} \times {\mathbf k} \cdot { \bm \sigma} h^+(q,k,\cos \theta)
  \right] \nonumber \\
&&+ i \epsilon^{abc} \tau^c \left[ g^-(q,k,\cos \theta) + i
  {\mathbf q} \times {\mathbf k} \cdot { \bm \sigma} h^-(q,k,\cos \theta)
  \right].
\end{eqnarray}
Notice that at the same order there is also a disconnected kernel,
corresponding to the diagram shown in Fig.~\ref{fig:pindisconnected}.
Analogously to what was done in the simple model of the previous
section, the disconnected contribution to the scattering kernel can be
absorbed by a local counterterm $\delta^{\mathrm{ren}}_{1+1}$ in the
corresponding diagonal entry of the mass operator.

Thus the  corresponding Lippmann-Schwinger equation reads,
\begin{equation}
t^I_{\ell \pm}(q,k) = b^I_{\ell \pm}(q,k) + \int \frac{\omega_{\mathbf p}
  p^2 dp}{(2 \pi)^2}\frac{b^I_{\ell \pm}(q,p) t^I_{\ell
  \pm}(p,k)}{\sqrt{s} -  \omega_{\mathbf p} - \omega_{\mathbf p}^\pi
   + i \epsilon},
\end{equation}
where 
\begin{equation}
b^I_{\ell \pm} (q,k) = \int_{-1}^1 d x \left\{ g^I(q,k,x) P_\ell (x) +
q k  h^I(q,k,x) \left[ P_{\ell \pm 1}(x) - x P_\ell (x) \right]
\right\},
\end{equation}
the functions $y^I$ ($y=g,h$) correspond to $y^{1/2}=y^+ + 2 y^-$
and $y^{3/2}=y^+-y^-$, and the subscript $\pm$ corresponds to total
angular momentum $J=\ell \pm 1/2$.
At the first order of the low-energy expansion, we can 
neglect the contributions of $\delta^{\mathrm{ren}}_1$.
The equation is then solved for the off-shell amplitude $t^I_{\ell
  \pm}(q,k)$ by discretizing the energy domain of
integration, for different choices of the c.o.m. energy
$\sqrt{s}$. 
Phaseshifts are then calculated by putting the particles
on energy shell,
\begin{equation}
{\mathrm{e}}^{2 i \delta^I_{\ell \pm}(E)} = 1 -2 \pi i \frac{E^2 p}{4
  ( 2 \pi)^2 } 
t^I_{\ell \pm}(p,p) = 1 + 2 i p f^I_{\ell \pm}(p),
\end{equation}
where $p$ is the c.o.m. momentum, $E=\sqrt{m_N^2 + p^2} +
\sqrt{M_\pi^2 + p^2}$. 
In the non-relativistic limit, realized as $m_N \to \infty$, our
results coincide with the ones obtained in the heavy baryon
formulation of ChPT \cite{fettes}, at the same chiral order. 
Although no complete fit procedure has been
performed, we have chosen the Weinberg-Tomozawa coupling constant so as to
reproduce the $I=1/2$ $S$-wave, while $g_A$ is fixed from the peripheral $NN$
phaseshifts (to be discussed later). This is repeated for three values of the
cutoff $\Lambda=300-400-500$~MeV.
\begin{figure}
\centerline{\includegraphics[width=16cm]{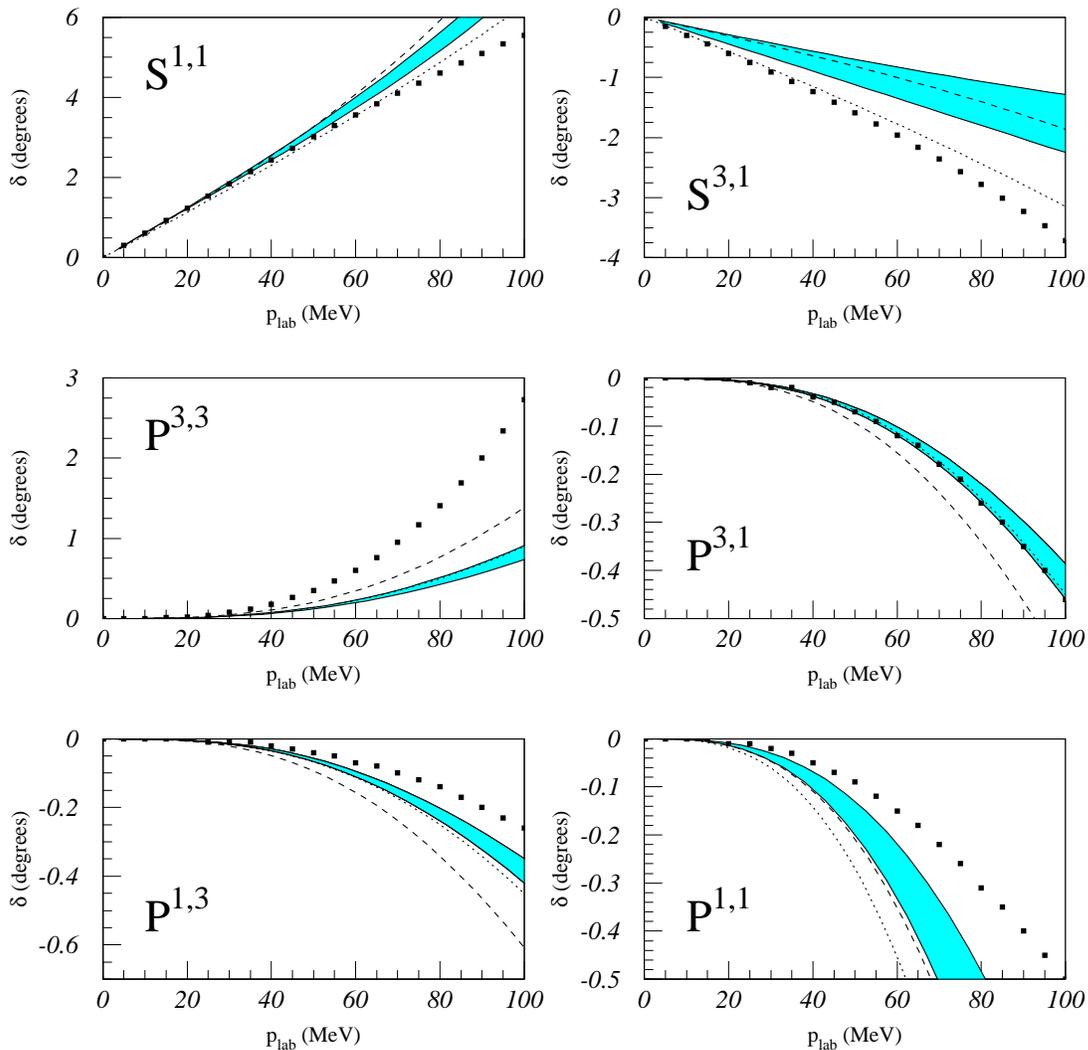}}
\caption{(Color online) $S$ and $P$ waves of $\pi N$ scattering as
  function of laboratory momentum. The (blue) bands are obtained by varying  the
   cutoff 
   $\Lambda$ between 300 and 500~MeV. Also shown are the experimental
  analysis\cite{pin} (squares), our results in the non-relativistic
  limit for $\Lambda=400$~MeV (dashed) and the three-level HBChPT
  result  (dotted).\label{fig:pin} } 
\end{figure}
In Fig.~\ref{fig:pin} the lowest partial waves are shown as function of
laboratory three-momentum, compared with the experimental analyses (squares)
\cite{pin}. 
The bands represent the uncertainties obtained by varying $\Lambda$ between
300 and 500~MeV.
However this is only a lower bound on the theoretical uncertainty: in
particular one should expect a larger uncertainty to come from the neglected
higher orders, especially from the contribution of the NLO couplings $c_i$'s
which are known to be  anomalously large because of the proximity of the
$\Delta$ \cite{meissner}. 
The description of $P$-wave phaseshifts is rather poor in the channel of the $\Delta$, and reflects the lack
of important physics not included in our leading order calculation. 
The dashed lines in Fig.~\ref{fig:pin} represent our result taken in the
non-relativistic limit ($m_N\to \infty$) for $\Lambda=400$~MeV, for which the
fit procedure has been repeated, and show that relativistic corrections to
this leading order calculation are sizeable in the case of $\pi N$ scattering.
The dotted lines represent
the tree-level HBChPT
calculation, which should be of comparable accuracy as ours.  One should
emphasize that the latter calculation is  
perturbative while ours takes into account, through the LS equation, the loops
due to rescattering. These loops in the (perturbative) HBChPT framework come
always together with subleading vertices, which absorb the cut-off
dependence. The renormalization of effective field theories at the
non-perturbative level is an intensively studied problem, currently  the
subject of 
controversies among experts \cite{workinggroupCD06}. The importance of
rescattering effects is especially apparent for the $I=3/2$ $S$-wave.

\section{Two-nucleon sector} \label{sec:nn}
In addition to ${\cal L}_{\pi N}$, for the two-nucleon sector one has to
consider, at the leading order of 
the low-energy expansion, also a contact interaction Lagrangian ${\cal
  L}_{NN}$ containing 4 nucleon fields and no derivatives. We therefore
consider in this section  ${\cal H}(x) = - {\cal L}_{\pi N} - {\cal L}_{NN}$
and discuss separately the one-pion exchange and the contact interaction.
\subsection{One-pion exchange}
The nucleon-nucleon wave equation can be written as
Eq.~(\ref{eq:eigenstates}). The two-particle component of the wave
function in momentum space will carry in this case spin ($r$,$r'$) and
isospin ($i$,$i'$) indices,  and in the center-of-mass system is written as
\begin{equation}
(2 \omega_{\mathbf k} +\delta^{\mathrm{ren}}_2({\mathbf k})) \phi_2 ({\mathbf k})_{r r',i i'}+ \int \omega_{\mathbf q}
\frac{d^3 {\mathbf q}}{(2\pi)^3} V({\mathbf k},{\mathbf q})^{s s', j
  j'}_{r r', i i'}  \phi_2 ({\mathbf q})_{s s',j j'} = \sqrt{s}
\phi_2 ({\mathbf k})_{r r',i i'} ,
\end{equation}
where
\begin{equation}
\phi_2({\mathbf k})_{r r', i i'} = \langle v=(1,{\mathbf 0});{\mathbf
  k}, r, i; - {\mathbf k},r',i' | \phi_2 \rangle.
\end{equation}
In analogy to Eq.~(\ref{eq:eigenwaves}), the Lippman-Schwinger equation will
again contain disconnected 
contributions of the kind of the function $A$ and connected
contributions of the kind of the function $B$,
\begin{equation}
V^{s s', j
  j'}_{r r', i i'} ({\mathbf k},{\mathbf q}) = \left[B^{s s', j
  j'}_{r r', i i'} ({\mathbf k},{\mathbf q}) + (2 \pi )^3 \delta^3 ({\mathbf k}
- {\mathbf q}) \delta_{r s} \delta_{r' s'} \delta_{ij} \delta_{i'j'}  A ({\mathbf k})\right]  - ({\mathbf q}
  \leftrightarrow - {\mathbf q}, s \leftrightarrow s', j \leftrightarrow j'),
\end{equation}
with
\begin{equation}
\begin{array}{rl}
B^{s s', j
  j'}_{r r', i i'}  ({\mathbf k},{\mathbf q} ) =&  \frac{g_A^2}{4
  F_{\pi}^2} \frac{1}{ 8\omega^\pi_{{\mathbf q} - {\mathbf 
  k}} \sqrt{\omega_{\mathbf q}^3 \omega_{\mathbf k}^3  }} \frac{f_A(2  \omega_{\mathbf
  k},\omega_{\mathbf q} + \omega^\pi_{{\mathbf q} - {\mathbf k}} + \omega_{\mathbf
  k}) f_A( \omega_{\mathbf k} + \omega^\pi_{{\mathbf q} - {\mathbf k}} +
  \omega_{\mathbf q},2  \omega_{\mathbf q}) }{\sqrt{s} -
  \omega_{\mathbf q} - \omega_{\mathbf k} - \omega^\pi_{{\mathbf q} -
  {\mathbf k}}}\times \\
&\times  \bar u ({\mathbf k},r) \gamma^\mu \gamma_5 u ({\mathbf q},s) \bar
  u(-{\mathbf k},r') \gamma^\nu \gamma_5 u(-{\mathbf q},s') p^\pi_\mu p^\pi_\nu
  {\bm \tau}_{ij}\cdot {\bm \tau}_{i'j'} \\
A({\mathbf k}) =& \frac{3 g_A^2}{4 F_{\pi}^2} \int \frac{d^3 {\mathbf q}}{(2
  \pi)^3} \frac{1}{ 8\omega_{\mathbf k}^2 \omega_{\mathbf q}
\omega^{\pi}_{{\mathbf q} - {\mathbf k}}}  \frac{|f_A(2  \omega_{\mathbf k},\omega_{\mathbf
  q} + \omega^\pi_{{\mathbf q} - {\mathbf k}} + \omega_{\mathbf k})|^2}{\sqrt{s} -
  \omega_{\mathbf q} - \omega_{\mathbf k} - \omega^\pi_{{\mathbf q} -
  {\mathbf k}}}\times \\
&\times \bar u
  ({\mathbf k},r) \gamma^\mu \gamma_5 u({\mathbf q},r ) \bar u({\mathbf q},r') \gamma^\nu
  \gamma_5 u ({\mathbf k},r') p^\pi_\mu p^\pi_\nu,
\end{array}
\end{equation}
where $f_A$ is the form factor entering the definition of the
interacting mass in Eq.~(\ref{eq:MIdef}) 
corresponding to the first operator in Eq.~(\ref{eq:Lpin}),
and we have denoted by $p^\pi$ the 4-momentum of the intermediate pion,
$p^\pi=(\omega^\pi_{{\mathbf q} - {\mathbf k}}, {\mathbf q} - {\mathbf
      k})$.
The disconnected contribution can be absorbed by the choice of the
counterterm $\delta^{\mathrm{ren}}_2({\mathbf k}) = - 2
\omega_{\mathbf k} A({\mathbf k})$. Counting the involved three momenta as small quantities of order
$O(p)$,  the pion mass $ M_\pi \sim O(p)$ and the energy denominators
$\sim O(p)$ we have that $\delta^{\mathrm{ren}}_2  \sim O(p^3)$.
The remaining connected kernel  $B$ (matrix in spin and isospin) reads
explicitly, 
\begin{equation}
  B({\mathbf q},{\mathbf k})_{r r', i i'}^{s s', j j'}   
=B_0({\mathbf q},{\mathbf k} ) \left[ ({\bm \sigma}\cdot {\mathbf
    p_1})_{r}^{s} ({ 
  \bm \sigma}\cdot {\mathbf p_1})_{r'}^{s'} - ({\bm \sigma}\cdot {\mathbf
  p_2})_{r}^{s} ({ 
  \bm \sigma}\cdot {\mathbf p_2})_{r'}^{s'} \right]
\left[ 2 \delta_i^{j'} \delta_{i'}^{j} - \delta_{i}^{j}
  \delta_{i'}^{j'} \right] 
 - \left\{{\mathbf k} \leftrightarrow -{\mathbf k} ; s \leftrightarrow
s' ; j \leftrightarrow j' \right\},
\end{equation}
with the vectors ${\mathbf p_1}$ and ${\mathbf p_2}$ defined by
\begin{eqnarray} \label{eq:p1p2}
{\mathbf p_1} &=& {\mathbf q} - {\mathbf k}  - \frac{{\mathbf k}^2 {\mathbf q} - {\mathbf q}^2 {\mathbf
  k}}{\left[\omega_{\mathbf k} + m_N \right] 
  \left[\omega_{\mathbf k} + m_N \right]} , \nonumber
\\
{\mathbf p_2} &=& \omega^\pi_{{\mathbf q}-{\mathbf k}} \left[ \frac{{\mathbf k}}{\omega_{\mathbf k} +
      m_N} + \frac{{\mathbf q}}{\omega_{\mathbf q} +
      m_N} \right],
\end{eqnarray}
and the scalar kernel
\begin{equation}
B_0({\mathbf q},{\mathbf k} ) 
=
\frac{g_A^2}{4 F_\pi^2} \frac{1}{  8
  \omega^\pi_{{\mathbf q}-{\mathbf k}} \sqrt{ \omega_{\mathbf q}^3 \omega_{\mathbf k}^3}} \left[\omega_{\mathbf q} + m_N \right]
  \left[\omega_{\mathbf k} + m_N \right]
\frac{f_A(2  \omega_{\mathbf k},\omega_{\mathbf q} +
  \omega^\pi_{{\mathbf q}-{\mathbf k}} + \omega_{\mathbf k})
  f_A(\omega_{\mathbf k} + \omega^\pi_{{\mathbf q}-{\mathbf k}} + \omega_{\mathbf
  q},2  \omega_{\mathbf q})  }{\sqrt{s} -
  \omega_{\mathbf q} - \omega_{\mathbf k} - 
  \omega^\pi_{{\mathbf q} - {\mathbf k}}}.
\end{equation}

\subsection{Contact interactions}
The most general chirally invariant leading order $NN$ Lagrangian can be
written, after using Fierz reordering, as the sum of two terms
\begin{equation}
{\cal L}_{NN} = -\frac{1}{2} C_S \bar \psi \psi \bar \psi \psi +
\frac{1}{2} C_T \bar \psi \gamma^\mu \gamma^5 \psi \bar \psi \gamma_\mu \gamma^5 \psi,
\end{equation}
where the notation for the coupling constants has been chosen so as to
conform with the usual ones in the non relativistic expansion. 
The two vertices are diagonal in the Fock space basis that we have
chosen,
\begin{equation} \label{eq:coupled}
\left( \begin{array}{cc} D_2 + \delta^{\mathrm{ren}}_2 + C_2 & g K^\dagger \\
g K & D_{2+1} + C_{2+1} \end{array} \right) \left( \begin{array}{c} \phi_2
  \\ \phi_{2+1} \end{array} \right) = \sqrt{s}  \left( \begin{array}{c} \phi_2
  \\ \phi_{2+1} \end{array} \right),
\end{equation}
where we have denoted by $C_2$ and $C_{2+1}$ the contribution of the new
interaction vertices to the interacting mass operator in the
two-particle and three-particle subspace. According to the recipe,
they contain in addition an associated structure function $f$.
The contribution of $C_{2+1}$ to the wave equation is of
higher chiral order, therefore in our leading order calculation,
we can neglect it 
and only include the operator $C_2$ in the analysis. With this
understanding the scattering equation projected in the two-particle
subspace becomes 
\begin{equation}
 D_2  \phi_2 + C_2 \phi_2 + \left. g^2 K^\dagger \left[ \sqrt{s} - D_{2+1} \right]
g K \right|_{\mathrm{conn}} \phi_2 = \sqrt{s} \phi_2.
\end{equation}
The new vertices yield connected  contributions to the wave
equation.
Let us denote this connected kernel, analogous to  $B({\mathbf k},{\mathbf q})_{r
  r', i i'}^{s 
  s', j j'}$ in the one-pion exchange potential, with the letter $C$. Its
  operatorial expression 
  reads  
\begin{eqnarray}
C({\mathbf k},{\mathbf q}) &=&
C_S \left\{ \left[ \sqrt{( \omega_{\mathbf k} + m_N) (\omega_{\mathbf q} + m_N)} - \frac{{\mathbf q}
    \cdot {\mathbf k}}{\sqrt{( \omega_{\mathbf k} + m_N) (\omega_{\mathbf q} + m_N)}}
\right]^2 \right. \nonumber \\
&&+ i \left[ 1 - \frac{ {\mathbf q}
    \cdot {\mathbf k}}{( \omega_{\mathbf k} + m_N) (\omega_{\mathbf q} + m_N)}\right] ({\mathbf
  k} \times {\mathbf q}) \cdot ({ \bm \sigma_1} + { \bm \sigma_2})\nonumber \\
&& \left.  - \frac{ ( {\mathbf
  k} \times {\mathbf q}) \cdot {\bm \sigma_1}  ( {\mathbf
  k} \times {\mathbf q}) \cdot {\bm \sigma_2}  }{( \omega_{\mathbf k} + m_N)
(\omega_{\mathbf q} + m_N)} \right\} \frac{f_S(2 \omega_{\mathbf q} , 2
\omega_{\mathbf k})}{8 \sqrt{\omega_{\mathbf q}^3 \omega_{\mathbf k}^3}}  \nonumber \\
&&+ C_T \left\{ \left[ ( \omega_{\mathbf k} + m_N) (\omega_{\mathbf q} + m_N) + \frac{({\mathbf q}
    \cdot {\mathbf k})^2}{( \omega_{\mathbf k} + m_N) (\omega_{\mathbf q} + m_N)}
\right] {\bm \sigma_1} \cdot {\bm \sigma_2}  \right. \nonumber \\
&& + \left[ 1 - \frac{ {\mathbf q}
    \cdot {\mathbf k} }{( \omega_{\mathbf k} + m_N) (\omega_{\mathbf q} + m_N)} \right] ({\mathbf k}
  + {\mathbf q}) \cdot {\bm \sigma_1} ({\mathbf k}
  + {\mathbf q}) \cdot {\bm \sigma_2}   \nonumber \\
&& + \left[ \frac{(\omega_{\mathbf k} - \omega_{\mathbf q}) ( \omega_{\mathbf k} + m_N) +
    {\mathbf k}^2 + {\mathbf k} \cdot {\mathbf q}}{( \omega_{\mathbf k} + m_N) (\omega_{\mathbf q}
    + m_N)} \right] {\mathbf q} \cdot { \bm \sigma_1} {\mathbf q} \cdot { \bm \sigma_2}
\nonumber \\
&& + \left[ \frac{(\omega_{\mathbf q} - \omega_{\mathbf k}) ( \omega_{\mathbf q} + m_N) +
    {\mathbf q}^2 + {\mathbf k} \cdot {\mathbf q}}{( \omega_{\mathbf k} + m_N) (\omega_{\mathbf q}
    + m_N)} \right] {\mathbf k} \cdot {\bm \sigma_1} {\mathbf k} \cdot {\bm \sigma_2}
\nonumber \\
&&\left.+i \frac{({\mathbf q} \cdot {\mathbf k} ) \, ({\mathbf k} \times {\mathbf q}) \cdot (
    {\bm \sigma_1} 
  + {\bm \sigma_2})}{( \omega_{\mathbf k} + m_N) (\omega_{\mathbf q} + m_N)} -
\frac{{\mathbf q}^2 {\mathbf k}^2 - ( {\mathbf q} \cdot {\mathbf k} )^2}{( \omega_{\mathbf k} +
  m_N) (\omega_{\mathbf q} + m_N)} \right\} \frac{f_T(2 \omega_{\mathbf q} , 2 \omega_{\mathbf
k})}{8 \sqrt{\omega_{\mathbf q}^3 \omega_{\mathbf k}^3}} \nonumber \\
&& - \left\{{\mathbf q} \leftrightarrow -{\mathbf q} ; 1 \leftrightarrow
2 \right\},
\end{eqnarray}
where for the associated structure functions $f_S$ and $f_T$ we take
the same expression as in Eq.~(\ref{eq:cutoff}), with the choice
$\xi=$exp$[-(q^4 + k^4)/(2 \Lambda^4)]$. Such additional cutoff for
the contact interactions is needed to regulate the LS equation. It is
chosen, analogously to what is done e.g. in Ref.~\cite{egm2}, so that
the induced modifications are of higher order in the chiral counting
(actually more than needed for our leading order calculation). 
The wave equation now reads
\begin{equation}
\label{eq:waveeq2}
 \int \omega_{\mathbf q} \frac{d^3{\mathbf q}}{(2 \pi)^3} 
\left[ B({\mathbf  k},{\mathbf q})_{r r',i i'}^{s s', j j'}+ C({\mathbf
  k},{\mathbf q})_{r r',i i'}^{s s', j j'} \right] \phi_2( {\mathbf
  q})_{s s', j j'} 
=\left[ \sqrt{s} - 2 \omega_{\mathbf k} \right] \phi_2( {\mathbf
  k})_{r r', i i'}.
\end{equation}
In the chiral counting $B({\mathbf k},{\mathbf q}) \sim 
C({\mathbf k},{\mathbf q}) \sim O(1)$.
The wave equation can be recast in the form of a
Lippmann-Schwinger equation, from which phaseshifts are found by means of
standard numerical methods \cite{davies}.

As already mentioned in the previous section, we have fixed the coupling $g_A$
from the $^1I_6$ wave, and the contact terms $C_S$ and $C_T$ from the $^3 S_1$
and $^1S_0$ scattering lenghts, requiring that
\begin{equation}
a( ^3 S_1) = 5.4~{\mathrm{fm}}, \quad a( ^1 S_0) = -24~{\mathrm{fm}}.
\end{equation}
We have repeated this procedure for three values of the cutoff
$\Lambda=300,400,500$~MeV.  
\begin{figure}
\centerline{\includegraphics[width=16cm]{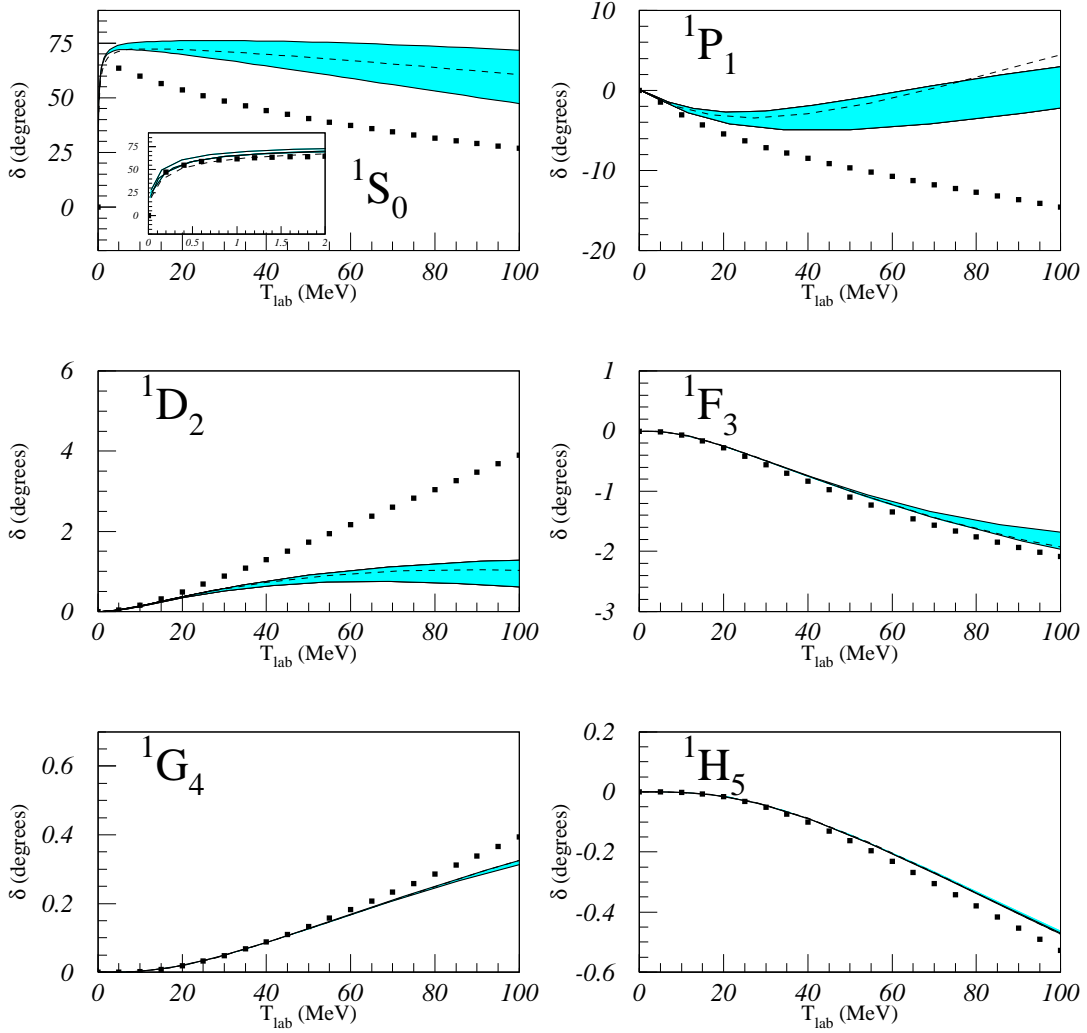}}
\caption{(Color online) Spin singlet $NN$ phaseshifts as function of
  laboratory kinetic energy. The (blue) bands show the effect of varying the cutoff
  $\Lambda$ between 300 and 500~MeV. The squares represent the Nijmegen
  phaseshifts \cite{nijmegen}. Also shown in dashed are the results in the
  nonrelativistic limit ($m_N \to \infty$), computed with
  $\Lambda=400$~MeV. \label{fig:nn1}}
\end{figure}
\begin{figure}
\centerline{\includegraphics[width=16cm]{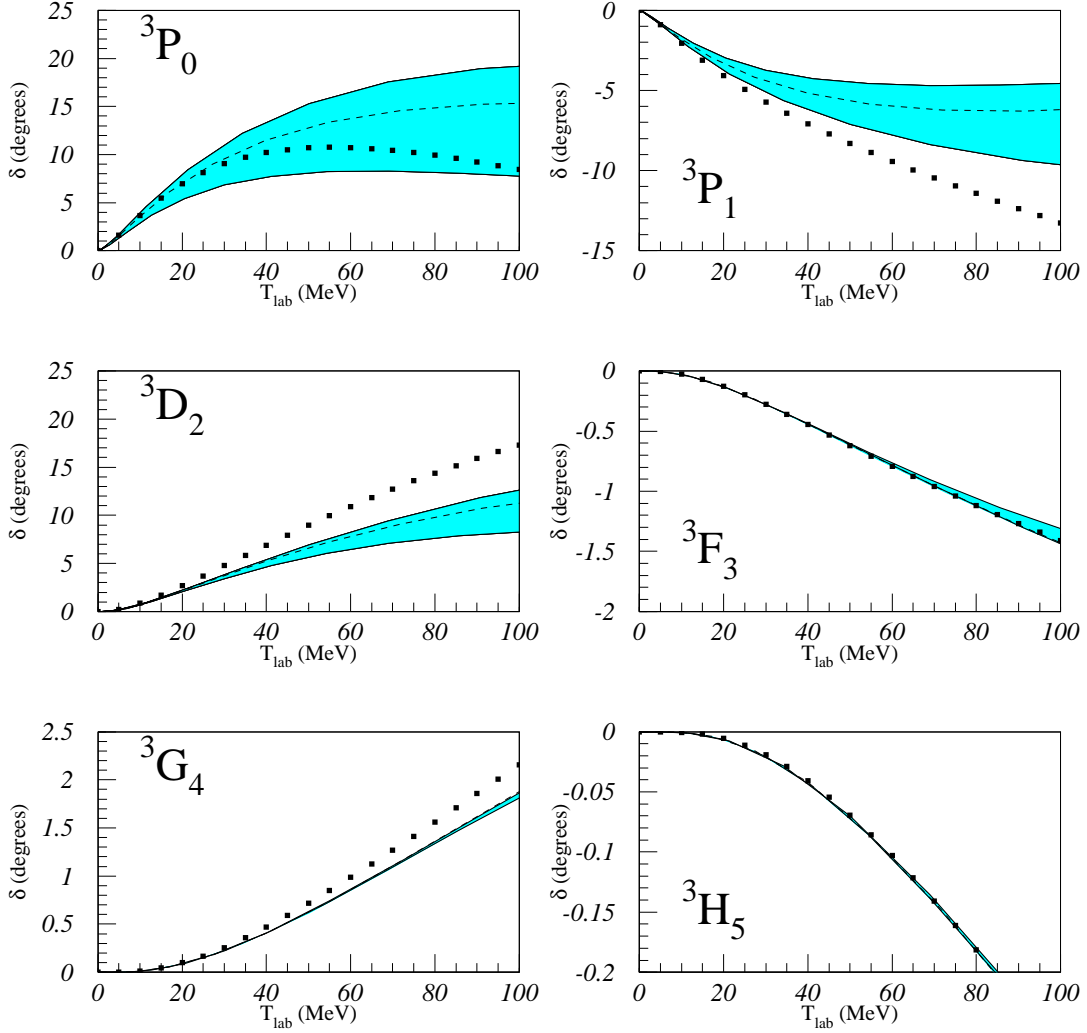}}
\caption{ (Color online) Spin triplet $NN$ phaseshifts in the uncoupled
  channels ($L=J$). The legend is as in Fig.~\ref{fig:nn1}. \label{fig:nn2}}
\end{figure}
\begin{figure}
\centerline{\includegraphics[width=16cm]{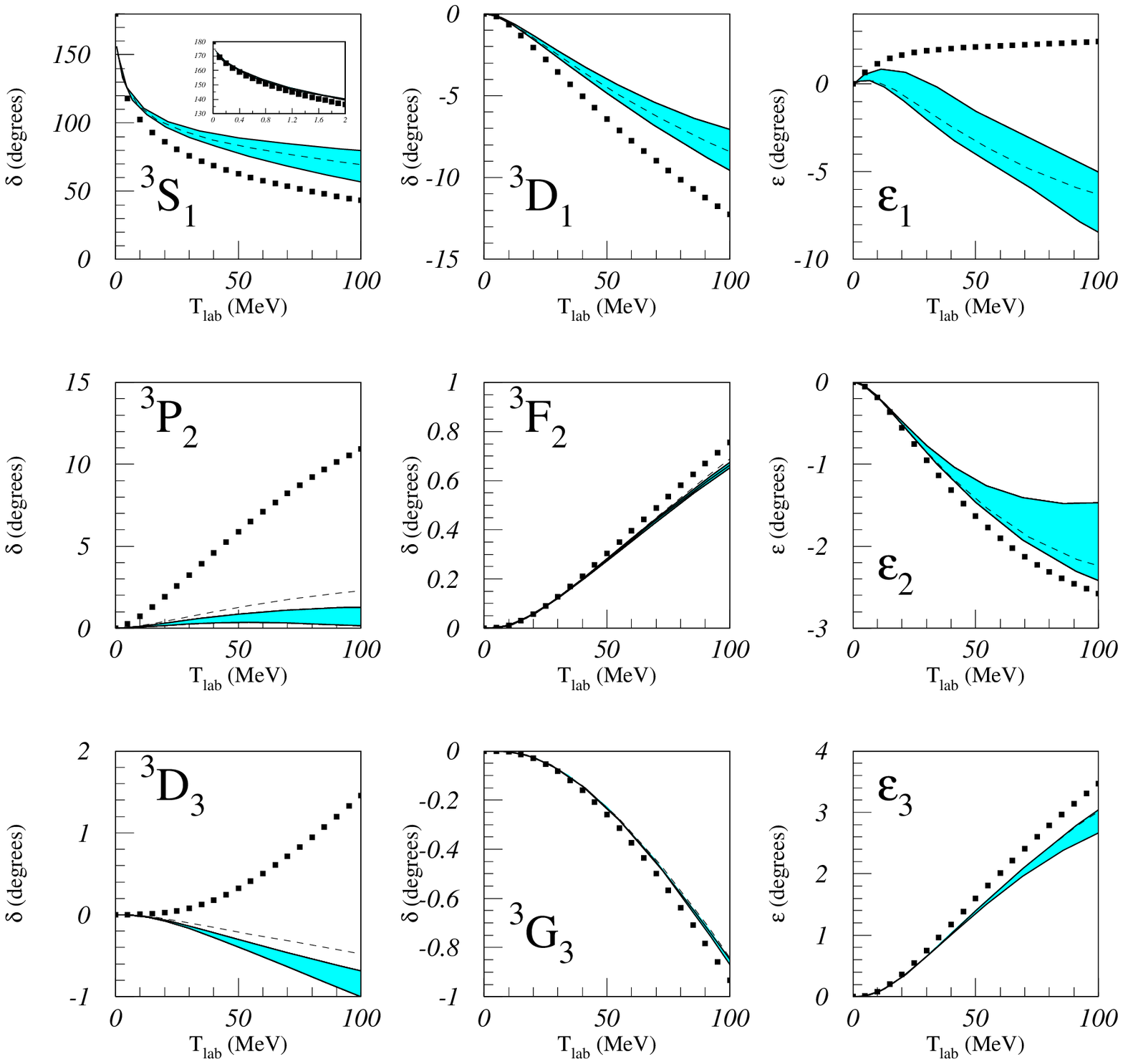}}
\caption{(Color online) Spin triplet  $NN$ phaseshifts and mixing angles in
  the coupled channels ($L=J\pm1$). The legend is as in Fig.~\ref{fig:nn1}. \label{fig:nn3}}
\end{figure}
\begin{figure}
\centerline{\includegraphics[width=16cm]{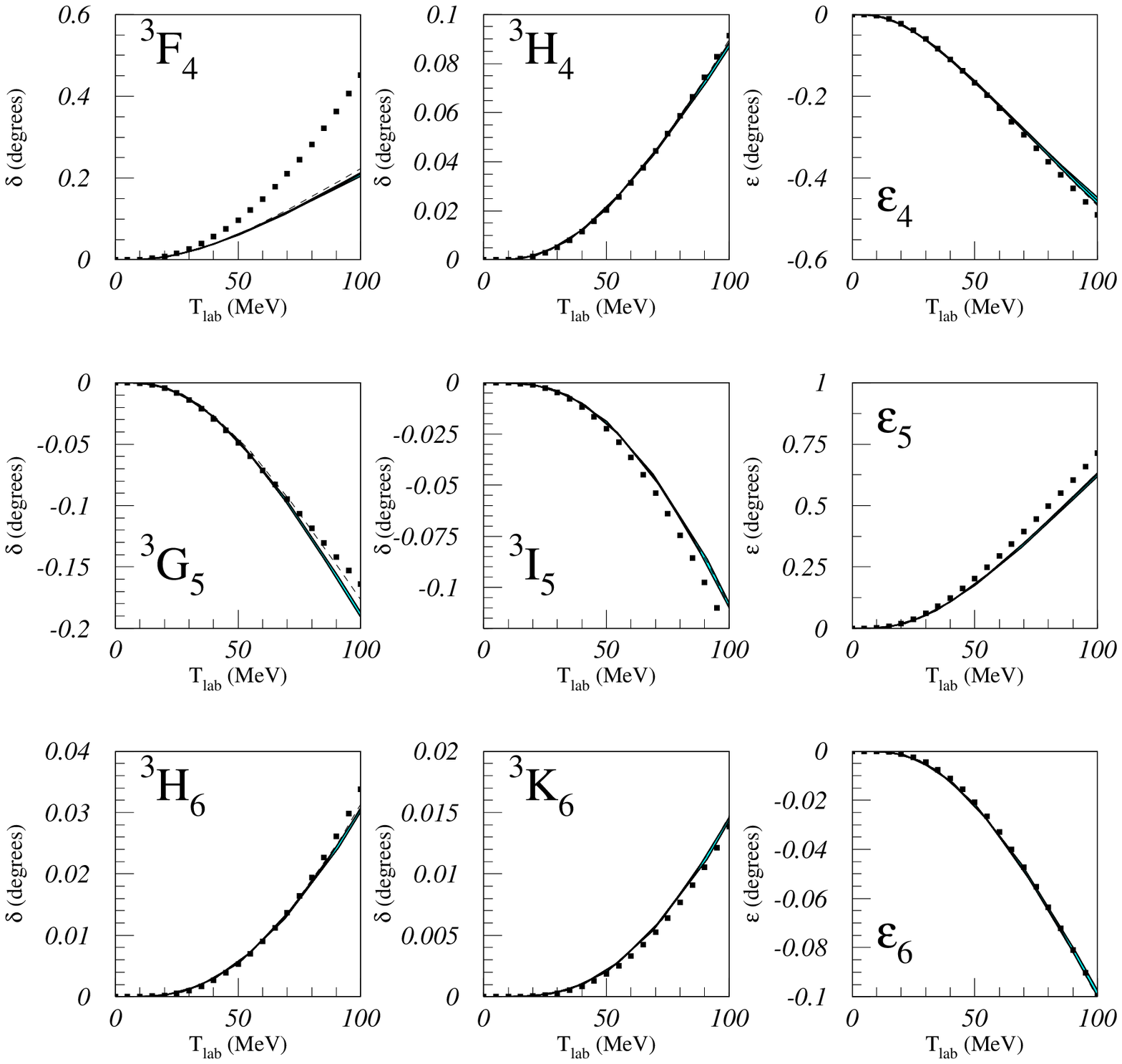}}
\caption{(Color online) Spin triplet  $NN$ phaseshifts and mixing angles in
  the coupled channels ($L=J\pm1$). The legend is as in Fig.~\ref{fig:nn1}. \label{fig:nn4}}
\end{figure}
The results are shown in
Figs.~\ref{fig:nn1}-\ref{fig:nn4}. The shaded bands represent the
  variations with $\Lambda$ and can be considered as the
  intrinsic theoretical uncertainty of our calculation. Also plotted, as
  dashed lines, is the result corresponding to $\Lambda=400$~MeV in the
  non-relativistic limit, realized as $m_N \to \infty$. The differences
  make it possible  to quantify the size of the relativistic corrections
  included in our 
  scheme: they are always smaller than the effect of varying $\Lambda$,
  therefore, at least for the leading order of the chiral expansion,
  relativistic corrections are smaller than the neglected chiral orders.
 The peripheral waves are well described by the one-pion
  exchange, but the agreement with data is poorer for the lowest
  waves. Clearly some important physics is missing in our leading order
  calculation (see e.g. the discussion of Ref.~\cite{egm2}). This  is the
  case, for instance, of the $^3D_3$ wave: it is well 
  known that two-pion exchange is very important for the description of this
  wave, and we do not include this process, not even implicitly,
  via  subleading low-energy constants. 

\subsection{The deuteron}
\begin{figure}
\centerline{\includegraphics[width=10cm]{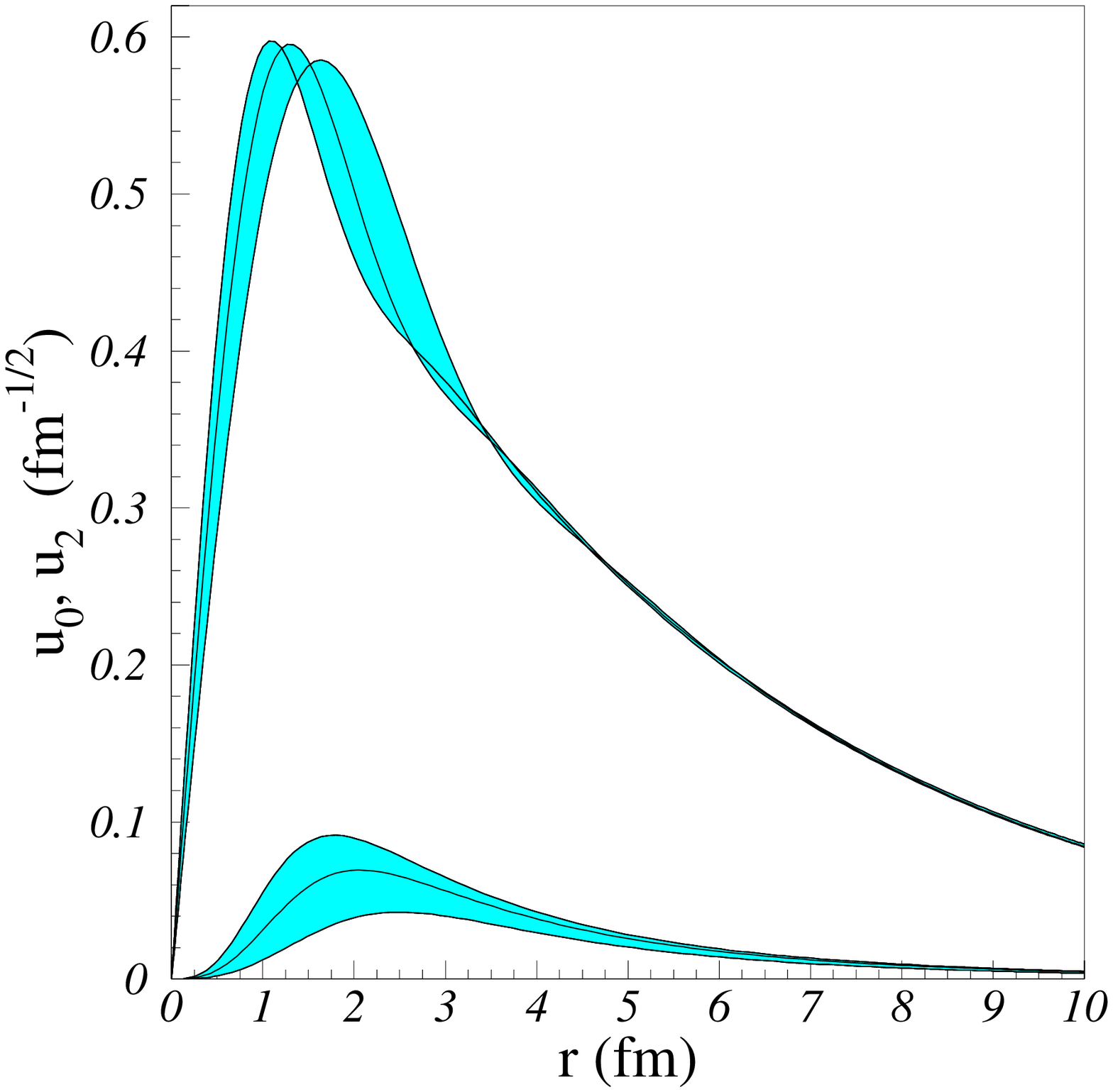}}
\caption{(Color online) Configuration-space wave function of the S-wave (upper curves) and
  D-wave (lower curves) components of the deuteron. The (blue) bands show the effect
  of varying the cutoff between 300 and 500~MeV.\label{fig:deuteron}}
\end{figure}

With all the parameters fixed by the scattering, for the chosen values of the
cutoff $\Lambda$, we have computed the momentum space bound state mass and
wave function $\phi_2^D({\mathbf k})$ in the $^3S_1$-$^3D_1$ channel. The
configuration-space wave functions $u_0$ and $u_2$ are found by Fourier transformation
(cfr. Ref.~\cite{machleidt}).  Some bound-state observables are shown in
Table~\ref{tab:deuteron}. 
\begin{table}
\begin{tabular}{l c c c c }
\hline 
 & $\Lambda=300$~MeV  & $\Lambda=400$~MeV  & $\Lambda=500$~MeV & Empirical
 (from Ref.~\cite{machleidt})\\
\hline
Binding energy $B_d$ (MeV) & 2.01 & 1.93 & 1.99 & 2.224575(9)\\
Asymptotic D/S state $\eta$ & 0.015& 0.018& 0.020 & 0.0256(4) \\
Matter radius $r_d$ (fm) & 1.86 & 1.85 & 1.83 & 1.971(6)\\ 
Quadrupole moment $Q_d$ (fm$^2$) & 0.12 & 0.16 & 0.18 & 0.2859(3) \\
D-state probability $P_D$ (\%) & 0.5 & 1.2 & 1.9 \\
\hline
\end{tabular}
\caption{\label{tab:deuteron} Bound state observables for the chosen values of
  the cutoff $\Lambda$. }
\end{table}
In Fig.~\ref{fig:deuteron} we show the configuration-space deuteron wave
functions, in which as before the bands represent the variations with
$\Lambda$ between 300 and 500~MeV.

\section{Concluding remarks}
In this paper we have considered the proposal of Ref.~\cite{klink00c} of a
point-form formulation of relativistic quantum mechanics consisting
in the construction of a Bakamjian-Thomas interacting mass operator for the
few-nucleon system. Taking the most general chirally invariant Lagrangian
describing pions and nucleons, we have implemented the above construction in
the 
framework of an effective theory, where the needed truncation of the Fock
space is justified by a systematic low-energy power counting. This in turn
relies on the restrictions that the underlying chiral symmetry imposes on the
interaction vertices. 
By introducing a different normalization for the structure functions
$f$, associated with 
each coupling of the Lagrangian, compared to the original proposal of
Ref.~\cite{klink00c}, we obtain the correct matching with the
point-form quantized field theory and show that the renormalization
procedure is consistent with the cluster decomposition principle.

We have performed a complete leading order analysis of the $\pi N$ and $NN$
systems by solving numerically the eigenvalue equations originating by the
diagonalization of the mass operator, and obtained a rather coherent picture of
the low-energy phaseshifts. Of course, especially in the $\pi N$ system,
important physics, such as the effect of the $\Delta$ resonance, is missed at
leading order, and the examination of higher orders is mandatory for a better
quantitative description. By comparison with the non-relativistic limit of our
framework we have assessed the size of relativistic corrections included in
our scheme, and shown that in the $NN$ sector they are always smaller than
the accuracy of our leading order calculation: NLO chiral corrections
are larger than our ``all-order'' relativistic corrections. Still, it
could well be, provided the chiral expansion is convergent enough,
that NNLO chiral corrections be smaller than our all-order
relativistic corrections. This can only be checked by actual
calculation. Moreover, the importance of relativistic
corrections can depend on the observables: in Ref.~\cite{miller} they
are found to be surprisingly large for the neutron-deuteron $A_y$. 

As a possible development then the next order can be analyzed, by 
including  the 2-pion states and taking into account the next-to-leading
order terms in ${\cal L}$. By examining the off-diagonal matrix elements of the mass operator,  our framework allows  to
incorporate in a natural way the pion production channel  whose understanding from the
field-theoretical point of view is only recently starting to emerge
\cite{lensky}. 
Furthermore, once the Lagrangian is fixed in the two-nucleon
sector, the $3N$ sector could be investigated. The point-form
formulation, in which the boost operators have only a kinematical
character, would allow to easily account for such relativistic
effects as the boosting of the wave function of an interacting
2-nucleon subsystem, although  the requirements of the cluster
decomposition principle might be less simple to fulfill in this case.

\end{document}